\documentclass[prd,nofootinbib,reprint,twocolumn,showpacs,superscriptaddress,floatfix]{revtex4-1}
\pdfoutput=1 
\usepackage{bm, color} 
\usepackage{amssymb,amsfonts,slashed,amsthm,amsmath, graphicx, soul, empheq, cancel, hyperref}
\usepackage[caption=false]{subfig}

\begin{document}

\newcommand{\vev}[1]{ \left\langle {#1} \right\rangle }
\newcommand{\bra}[1]{ \langle {#1} | }
\newcommand{\ket}[1]{ | {#1} \rangle }
\newcommand{\eV}{ \ {\rm eV} }
\newcommand{\KeV}{ \ {\rm keV} }
\newcommand{\MeV}{\  {\rm MeV} }
\newcommand{\GeV}{\  {\rm GeV} }
\newcommand{\TeV}{\  {\rm TeV} }
\newcommand{\1}{\mbox{1}\hspace{-0.25em}\mbox{l}}
\newcommand{\Red}[1]{{\color{red} {#1}}}

\newcommand{\lmk}{\left(}  
\newcommand{\rmk}{\right)}
\newcommand{\lkk}{\left[}  
\newcommand{\rkk}{\right]}
\newcommand{\lhk}{\left \{ }  
\newcommand{\rhk}{\right \} }
\newcommand{\del}{\partial}  
\newcommand{\la}{\left\langle} 
\newcommand{\ra}{\right\rangle}
\newcommand{\half}{\frac{1}{2}}

\newcommand{\bea}{\begin{array}}
\newcommand{\eea}{\end{array}}
\newcommand{\beq}{\begin{eqnarray}}
\newcommand{\eeq}{\end{eqnarray}}
\newcommand{\eq}[1]{Eq.~(\ref{#1})}

\newcommand{\dd}{\mathrm{d}}
\newcommand{\Mpl}{M_{\rm Pl}}
\newcommand{\mg}{m_{3/2}}
\newcommand{\abs}[1]{\left\vert {#1} \right\vert}
\newcommand{\mphi}{m_{\phi}}
\newcommand{\Hz}{\ {\rm Hz}}
\newcommand{\for}{\quad \text{for }}
\newcommand{\Min}{\text{Min}}
\newcommand{\Max}{\text{Max}}
\newcommand{\Kahler}{K\"{a}hler }
\newcommand{\cphi}{\varphi}
\newcommand{\Tr}{\text{Tr}}
\newcommand{\diag}{{\rm diag}}

\newcommand{\SUf}{SU(3)_{\rm f}}
\newcommand{\Upq}{U(1)_{\rm PQ}}
\newcommand{\Zpq}{Z^{\rm PQ}_3}
\newcommand{\Cpq}{C_{\rm PQ}}
\newcommand{\ubar}{u^c}
\newcommand{\dbar}{d^c}
\newcommand{\ebar}{e^c}
\newcommand{\nubar}{\nu^c}
\newcommand{\Ndw}{N_{\rm DW}}
\newcommand{\Fpq}{F_{\rm PQ}}
\newcommand{\fpq}{v_{\rm PQ}}
\newcommand{\Br}{{\rm Br}}
\newcommand{\Lag}{\mathcal{L}}
\newcommand{\Lqcd}{\Lambda_{\rm QCD}}

\newcommand{\ji}{j_{\rm inf}} 
\newcommand{\jb}{j_{B-L}} 
\newcommand{\M}{M} 
\newcommand{\im}{{\rm Im} }
\newcommand{\re}{{\rm Re} }

\def\lrf#1#2{ \left(\frac{#1}{#2}\right)}
\def\lrfp#1#2#3{ \left(\frac{#1}{#2} \right)^{#3}}
\def\lrp#1#2{\left( #1 \right)^{#2}}
\def\REF#1{Ref.~\cite{#1}}
\def\SEC#1{Sec.~\ref{#1}}
\def\FIG#1{Fig.~\ref{#1}}
\def\EQ#1{Eq.~(\ref{#1})}
\def\EQS#1{Eqs.~(\ref{#1})}
\def\TEV#1{10^{#1}{\rm\,TeV}}
\def\GEV#1{10^{#1}{\rm\,GeV}}
\def\MEV#1{10^{#1}{\rm\,MeV}}
\def\KEV#1{10^{#1}{\rm\,keV}}
\def\blue#1{\textcolor{blue}{#1}}
\def\red#1{\textcolor{blue}{#1}}

\newcommand{\eff}{\Delta N_{\rm eff}}
\newcommand{\neff}{\Delta N_{\rm eff}}
\newcommand{\cc}{\Omega_\Lambda}
\newcommand{\Mpc}{\ {\rm Mpc}}
\newcommand{\Msolar}{M_\odot}

\def\sn#1{\textcolor{red}{#1}}
\def\SN#1{\textcolor{red}{[{\bf SN:} #1]}}
\def\my#1{\textcolor{blue}{#1}}
\def\MY#1{\textcolor{blue}{[{\bf MY:} #1]}}
\def\yn#1{\textcolor{magenta}{#1}}
\def\YN#1{\textcolor{magenta}{[{\bf YN:} #1]}}

%%%%%%%%%%%%%%%%%%%%%%%%%%%%%%%%%%%%%%%%%%%%%%%%%%%%%%%%%%%%%%%
%######################
\begin{flushright}
TU-1210
\end{flushright}
%######################

\title{
Dynamics of Superconformal Axion: Quality and Scalegenesis
}

\author{Shota Nakagawa$^{1,2,3}$,
Yuichiro Nakai$^{1,2}$,
Masaki Yamada$^{3,4}$,
and Yufei Zhang$^{1,2}$
\\*[10pt]
$^1${\it \normalsize Tsung-Dao Lee Institute, Shanghai Jiao Tong University, \\
520 Shengrong Road, Shanghai 201210, China} \\*[3pt]
$^2${\it \normalsize School of Physics and Astronomy, Shanghai Jiao Tong University, \\
800 Dongchuan Road, Shanghai 200240, China}\\*[3pt]
$^3${\it \normalsize Department of Physics, Tohoku University, Sendai, Miyagi 980-8578, Japan}\\*[3pt]
$^4${\it \normalsize Frontier Research Institute for Interdisciplinary Sciences, Tohoku University,\\ Sendai, Miyagi 980-8578, Japan} \\
}

\begin{abstract}
We explore a dynamical mechanism to realize the emergence of a global $U(1)_{\rm PQ}$ symmetry
and its spontaneous breaking at an intermediate scale for an axion solution to the strong CP problem.
Such a dynamics is provided by a new supersymmetric QCD near the middle of conformal window
that couples to fields spontaneously breaking the $U(1)_{\rm PQ}$ symmetry.
A large anomalous dimension of the $U(1)_{\rm PQ}$ breaking fields leads to
the suppression of explicit $U(1)_{\rm PQ}$-violating higher dimensional operators.
The $U(1)_{\rm PQ}$ breaking vacuum is generated
at a scale hierarchically smaller than the Planck scale by a non-perturbative effect.
The $U(1)_{\rm PQ}$ breaking drives the conformal breaking, and all the new quarks become massive.
The axion potential is generated by the ordinary color $SU(3)_C$ effect
as the $U(1)_{\rm PQ}$ symmetry is only anomalous under the $SU(3)_C$.
The saxion direction is stabilized by supersymmetry breaking
and cosmologically harmless. 
\end{abstract}

\maketitle
\flushbottom

%%%%%%%%%%%%%%%%%%%%%%%%%%%%%%%%%%%%%%%%%%%%%%%%%%%%%%%%%%%
%%%%%%%%%%%%%%%%%%%%%%%%%%%%%%%%%%%%%%%%%%%%%%%%%%%%%%%%%%%
\section{Introduction
\label{introduction}}

The Peccei-Quinn (PQ) mechanism
\cite{Peccei:1977hh} is one of the most popular solutions to the strong CP problem
that asks why quantum chromodynamics (QCD) does not violate CP symmetry
\cite{Jackiw:1976pf,Callan:1976je}.
The PQ mechanism introduces an anomalous global $U(1)_{\rm PQ}$ symmetry that is spontaneously broken,
resulting in the existence of a pseudo-Nambu-Goldstone boson (pNGB) called {\it an axion}
\cite{Weinberg:1977ma,Wilczek:1977pj} (for reviews, see e.g. refs.~\cite{Kim:2008hd,DiLuzio:2020wdo}).
The axion dynamically cancels the CP-violating $\theta$-angle, thus solving the strong CP problem.
The $U(1)_{\rm PQ}$ breaking scale is hierarchically smaller than the Planck scale, and is given by $10^8 \, {\rm GeV} \lesssim f_a \lesssim 10^{12} \, \rm GeV$, known as the axion window. 
While the lower bound is determined by the stellar cooling process \cite{Mayle:1987as,Raffelt:1987yt,Turner:1987by,Chang:2018rso,Carenza:2019pxu,Leinson:2014ioa,Hamaguchi:2018oqw,Leinson:2019cqv,Buschmann:2021juv}, the upper one comes from the argument that the axion saturates the observed abundance of dark matter via the misalignment mechanism \cite{Preskill:1982cy,Abbott:1982af,Dine:1982ah} for a natural misalignment angle.
If the spontaneous $U(1)_{\rm PQ}$ breaking is driven by some perturbative dynamics,
we encounter a hierarchy problem as in the case of the electroweak symmetry breaking.
Various approaches to this axion hierarchy problem have been explored. 
They include composite axion models
\cite{Kim:1984pt,Choi:1985cb,Randall:1992ut,Izawa:2002qk,Yamada:2015waa,Redi:2016esr,DiLuzio:2017tjx,Lillard:2017cwx,Lillard:2018fdt,Gavela:2018paw,Lee:2018yak,Yamada:2021uze,Ishida:2021avk,Contino:2021ayn},
warped extra dimension models
\cite{Flacke:2006ad,Cox:2019rro,Bonnefoy:2020llz,Lee:2021slp}
and supersymmetric models
\cite{Barger:2004sf,Hall:2014vga,Feldstein:2012bu,Harigaya:2015soa,Harigaya:2017dgd}.

While the $U(1)_{\rm PQ}$ breaking scale $f_a$ is much smaller than the Planck scale,
at the same time, it is hierarchically larger than the electroweak scale and QCD scale.
Since the axion potential generated by QCD effects is suppressed by $f_a$,
it is fragile and a tiny explicit $U(1)_{\rm PQ}$-violating term can easily spoil the solution to the strong CP problem
in light of a stringent upper limit on the effective $\theta$-angle, $|\bar{\theta}|\lesssim 10^{-10}$, given by
non-observation of the neutron electric dipole moment
\cite{Baker:2006ts,Pendlebury:2015lrz}.
In fact, it has been argued that quantum gravity effects do not respect a global symmetry,
and we generally anticipate explicit $U(1)_{\rm PQ}$-violating higher dimensional operators
suppressed by appropriate powers of the Planck scale, which reintroduces the strong CP problem
\cite{Dine:1986bg, Barr:1992qq, Kamionkowski:1992mf, Kamionkowski:1992ax, Holman:1992us, Kallosh:1995hi, Carpenter:2009zs, Carpenter:2009sw}.
This issue has been known as {\it the axion quality problem}.
In addition to the composite axion models and warped extra dimension models,
for example, the introduction of a gauge symmetry to protect the $U(1)_{\rm PQ}$ symmetry
\cite{Cheng:2001ys,Harigaya:2013vja,Fukuda:2017ylt,Fukuda:2018oco,Ibe:2018hir,Choi:2020vgb,Yin:2020dfn,Chen:2021haa}
or a non-minimal coupling to gravity suppressing a wormhole contribution
\cite{Hamaguchi:2021mmt} may address the problem.

In the present paper, we pursue a possibility to simultaneously solve the axion hierarchy problem and the quality problem
by using a 4D superconformal dynamics.
The use of such dynamics to address the axion quality problem has been initiated in ref.~\cite{Nakai:2021nyf},
which introduces a new supersymmetric QCD in conformal window
that couples to fields spontaneously breaking the $U(1)_{\rm PQ}$ symmetry.
A large anomalous dimension of the $U(1)_{\rm PQ}$ breaking fields then 
suppresses explicit $U(1)_{\rm PQ}$-violating higher dimensional operators to realize a high-quality axion.
The spontaneous $U(1)_{\rm PQ}$ breaking
drives the conformal breaking, so that all the new (s)quarks become massive
and the remaining pure super Yang-Mills theory confines.
The model is constructed such that the $U(1)_{\rm PQ}$ symmetry is anomalous only under the ordinary color $SU(3)_C$, so that 
the axion potential is correctly generated by the $SU(3)_C$ effect.
In ref.~\cite{Nakai:2021nyf}, the spontaneous $U(1)_{\rm PQ}$ breaking is assumed and put by hand.
However, we here discuss that the model possesses an intrinsic mechanism to generate
a viable $U(1)_{\rm PQ}$ breaking vacuum and address the axion hierarchy problem.

Our mechanism to dynamically generate a $U(1)_{\rm PQ}$ breaking vacuum resembles
the holographically dual description of the Goldberger-Wise mechanism
\cite{Goldberger:1999uk} for radion stabilization in the Randall-Sundrum model
\cite{Randall:1999ee}.
That is, the model contains a small marginally relevant deformation in the superpotential,
$\Delta W = g \mathcal{O}_g$ where ${\rm Dim}[\mathcal{O}_g] = 3-\epsilon$ with $\epsilon \ll 1$.
The coupling $g$ receives a renormalization group (RG) evolution until it reaches a critical value $g_c$
at which the conformal invariance is lost.
Then, the conformal breaking scale is hierarchically smaller than a UV scale that the theory is defined.
In our model, a supersymmetric mass term of the $U(1)_{\rm PQ}$ breaking fields corresponds to such a deformation
$\Delta W$
and is balanced with a non-perturbative effect of the SQCD to generate a $U(1)_{\rm PQ}$ breaking vacuum.
According to the AdS/CFT correspondence
\cite{Maldacena:1997re,Gubser:1998bc,Witten:1998qj},
the idea of suppressing explicit $U(1)_{\rm PQ}$-violating operators by a superconformal dynamics
is similar to warped extra dimension models
\cite{Flacke:2006ad,Cox:2019rro,Bonnefoy:2020llz,Lee:2021slp},
where the axion hierarchy problem is also addressed due to a warp factor
with an appropriate mechanism of radion stabilization.
In this sense, our model is more minimal as the $U(1)_{\rm PQ}$ breaking fields
play a role in stabilizing the dilaton degree of freedom,
which is the 4D counterpart of the radion.

The rest of the present paper is organized as follows.
In Sec.~\ref{sec:model}, we present our model
and explain how a viable $U(1)_{\rm PQ}$ breaking vacuum is generated at a scale
hierarchically smaller than the Planck scale. 
Sec.~\ref{sec:CW} discusses the effect of supersymmetry (SUSY) breaking
and shows that the saxion direction can be properly stabilized.
Then, in Sec.~\ref{sec:PQquality}, we estimate the quality of the $U(1)_{\rm PQ}$ symmetry
that can be achieved in our model.
Sec.~\ref{sec:conclusion} is devoted to conclusions and discussions,
where we briefly discuss cosmological implications of our scenario.
App.~\ref{sec:axino} comments on the fermion components of the $U(1)_{\rm PQ}$ breaking fields.

%%%%%%%%%%%%%%%%%%%%%%%%%%%%%%%%%%%%%%%%%%%%%%%%%%%%%%%%%%%
%%%%%%%%%%%%%%%%%%%%%%%%%%%%%%%%%%%%%%%%%%%%%%%%%%%%%%%%%%%
\section{A superconformal axion model
\label{sec:model}}

Let us consider a supersymmetric $SU(N)$ gauge theory with $N_F$ vector-like fields, $Q$ and $\bar{Q}$,
which transform as the fundamental and anti-fundamental representations, respectively. 
We assume that the theory is in conformal window, $3N/2 < N_F < 3N$, and
has a nontrivial IR fixed point
\cite{Intriligator:2007cp}.
The number of flavors $N_F$ is taken to be an even number and 
parameterized as $N_F = (2 + \epsilon)N$, 
where we focus on a small $\epsilon$ whose lowest value is $\epsilon_{\rm min}=2/N$. 
We will see shortly that the size and sign of $\epsilon$ play an essential role in
getting a viable $U(1)_{\rm PQ}$ breaking vacuum. 
A pair of gauge-singlet chiral superfields, $\Phi$ and $\bar{\Phi}$, is introduced and has 
the following superpotential: 
\beq
W_Q = \lambda \Phi Q_m \bar{Q}_m + \bar{\lambda} \bar{\Phi} Q_k \bar{Q}_k \, . 
\label{WQ}
\eeq
Here, $\lambda$ and $\bar{\lambda}$ are dimensionless coupling constants, flowing into a nontrivial IR fixed point, and 
the subscripts $m$ $\in (1,2,\dots,N_F/2)$ and $k$ $\in (N_F/2+1,N_F/2+2,\dots,N_F)$ denote flavor indices. 
The superpotential respects $U(1)_{\rm PQ}\times SU(N_F/2)_1\times SU(N_F/2)_2$ flavor symmetry 
and $U(1)_R$ symmetry. 
The charge assignments are summarized in Table \ref{tab:charge}. 
We assume that a subgroup of the flavor symmetry is weakly gauged and identified as the ordinary color gauge group,
$SU(3)_{C}\subset {SU}(N_F/2)_{1}$. 
Then, the $U(1)_{\rm PQ}$ symmetry is anomalous under the $SU(3)_{C}$,
which implies that the PQ mechanism works after the spontaneous breaking,
as in the case of the KSVZ axion model~\cite{Kim:1979if,Shifman:1979if}. 
Note that $Z_N$ ($\subset {U}(1)_{\rm PQ}$) symmetry is not broken by the $SU(3)_C$ anomaly.
We impose this anomaly-free discrete symmetry to realize the $U(1)_{\rm PQ}$ symmetry at the renormalizable level,
but it also helps to suppress explicit $U(1)_{\rm PQ}$-violating higher dimensional operators on top of superconformal dynamics,
as will be shown later.

\begin{table}[t!]
\vspace{0mm}
\centering
\begin{tabular}{c|c|c|c|c|c|c}
& $Q_m$ & $\bar{Q}_m$ & $Q_k$ & $\bar{Q}_k$ & $\Phi$ & $\bar{\Phi}$\\
\hline
$SU(N)$ & $\square$ & $\bar{\square}$ & $\square$ & $\bar{\square}$ & ${\bm 1}$ & ${\bm 1}$ \\
$SU(N_F/2)_1$ & $\square$ & $\bar{\square}$ & ${\bm 1}$ & ${\bm 1}$ & ${\bm 1}$ & ${\bm 1}$\\
$SU(N_F/2)_2$ & ${\bm 1}$ & ${\bm 1}$ & $\square$ & $\bar{\square}$ & ${\bm 1}$ & ${\bm 1}$\\
$U(1)_{\rm PQ}$ ($\supset Z_N$) & $-1$ & 0 & 1 & 0 & 1 & $-1$ \\
$U(1)_{R}$ & $\frac{N_F-N}{N_F}$ & $\frac{N_F-N}{N_F}$ & $\frac{N_F-N}{N_F}$ & $\frac{N_F-N}{N_F}$ & $\frac{2N}{N_F}$ & $\frac{2N}{N_F}$ \\
\end{tabular}
\vspace{1mm}
\caption{The charge assignments in our model.}
\label{tab:charge}
\end{table}

A non-perturbative effect of the SQCD dynamically generates a superpotential for $\Phi$ and $\bar{\Phi}$.
Suppose they obtain nonzero vacuum expectation values (VEVs). 
The supermultiplets $Q_m, \bar{Q}_m$ and $Q_k, \bar{Q}_k$ then acquire masses of
$\lambda \Phi$ and $\bar{\lambda} \bar{\Phi}$, respectively. 
Since these quark supermultiplets are decoupled at the energy scale around their masses, 
the theory becomes a pure super Yang-Mills and shows gaugino condensation
at a slightly lower energy scale,
\beq
 &&W_{\rm gaugino} = N \Lambda_{\rm new}^3 \, ,\label{gaugino}
 \\[1ex]
 &&\Lambda_{\rm new}^3 = \lmk \lambda \Phi \rmk^{1+\epsilon/2} \lmk \bar{\lambda} \bar{\Phi} \rmk^{1+\epsilon/2}
 \Lambda'^{1-\epsilon} \, , \label{gaugino2}
\eeq
where $\Lambda'$ denotes the dynamical scale of the original theory with massless $Q$ and $\bar{Q}$, 
whereas 
$\Lambda_{\rm new}$ is the one of the pure super Yang-Mills which is determined by matching the dynamical scales
at the decoupling of $Q$ and $\bar{Q}$. 
We now introduce a supersymmetric mass term for $\Phi$ and $\bar{\Phi}$ as
\beq
W_\Phi = - M_\Phi \Phi \bar{\Phi} \, , 
\label{PhibarPhi} 
\eeq
with a mass parameter $M_\Phi$. 
This term explicitly breaks the $U(1)_R$ symmetry as well as the conformal symmetry, but
the scaling dimension of the operator at the fixed point is $3-3\epsilon/2$
(see the anomalous dimension of $\Phi$ and $\bar{\Phi}$ given below) and
only a small marginally relevant deformation for $\epsilon \ll 1$. 
It is naturally expected that the mass scales $M_\Phi, \Lambda'$ are not significantly smaller than a cutoff scale,
such as the Planck scale. 
The dynamically generated superpotential \eqref{gaugino} with Eq.~\eqref{gaugino2} and the mass term \eqref{PhibarPhi}
lead to the $F$-term potential for the scalar components of $\Phi$ and $\bar{\Phi}$,
\beq
V_F (\Phi, \bar{\Phi}) &=& 
\left| M_\Phi \Phi - \lmk1+\frac{\epsilon}{2}\rmk N \bar{\lambda} \Lambda'^{1-\epsilon} \lmk \lambda \Phi \rmk^{1+\epsilon/2} \lmk \bar{\lambda} \bar{\Phi} \rmk^{\epsilon/2} \right|^2 
\nonumber\\
&+& 
\left| M_\Phi \bar{\Phi} - \lmk1+\frac{\epsilon}{2}\rmk N \lambda \Lambda'^{1-\epsilon} \lmk \lambda \Phi \rmk^{\epsilon/2} \lmk \bar{\lambda} \bar{\Phi} \rmk^{1+\epsilon/2} \right|^2.\nonumber\\
&&
\label{VF}
\eeq
Here and hereafter, the scalar components for $\Phi$ and $\bar{\Phi}$ are represented by the same characters as the corresponding superfields.
The $F$-term potential has two degenerate minima: 
\beq
 \la \Phi \ra &=& \la \bar{\Phi} \ra =0 \, ,
\eeq
and
\beq
 \la \Phi \bar{\Phi} \ra &=& \frac{\Lambda'^2}{\lambda \bar{\lambda}} \lmk \frac{2}{2+\epsilon}\frac{M_\Phi}{N \lambda \bar{\lambda} \Lambda'} \rmk^{2/\epsilon}.
\label{flatdir}
\eeq
While the former vacuum is trivial, the latter one spontaneously breaks the $U(1)_{\rm PQ}$ symmetry. 
We assume that the vacuum of \eq{flatdir} is realized throughout the history of the Universe. 
Note that there is a modulus along with $\Phi \propto \bar{\Phi}^{-1}$, which corresponds to a saxion direction. 
Although the saxion has an exactly flat potential in the SUSY limit, SUSY breaking effects lift up the potential,
which will be discussed in Sec.~\ref{sec:CW}.

It is important to note that \eq{flatdir} is not directly identified as the $U(1)_{\rm PQ}$ breaking scale
squared because the kinetic terms of $\Phi$ and $\bar{\Phi}$ experience a significant wavefunction renormalization
in the conformal regime. 
The wavefunction renormalization factors for $\Phi, \bar{\Phi}$ and $Q, \bar{Q}$ are respectively given by
\cite{Nakai:2021nyf}
\beq
Z_\Phi &=& \left(\frac{M_c}{\Lambda}\right)^{-\gamma_\Phi},\\[1ex]
Z_Q &=& \left(\frac{M_c}{\Lambda}\right)^{-\gamma_Q},
\label{wfrenorm}
\eeq
where $\gamma_\Phi\equiv (2-2\epsilon)/(2+\epsilon)$ 
is the anomalous dimension of $\Phi$ and $\bar{\Phi}$ and $\gamma_Q \equiv (\epsilon-1)/(2+\epsilon)$ 
is that of $Q_{m(k)}$ and $\bar{Q}_{m(k)}$ at the fixed point. 
Here, $\Lambda$ and $M_c$ represent scales at which the theory enters into and
exits from the conformal regime, respectively. 
We denote canonically normalized $U(1)_{\rm PQ}$ breaking fields and vector-like quarks as
$\hat{\Phi}$, $\hat{\bar{\Phi}}$, $\hat{Q}$ and $\hat{\bar{Q}}$:
\beq
 \hat{\Phi} = 
 \sqrt{Z_\Phi} 
 &&~~
 \hat{\bar{\Phi}} =  
 \sqrt{Z_\Phi} 
 \bar{\Phi} \, ,
\label{eq:hatPhi}
 \\[1ex]
 \hat{Q}_{m(k)} =  
 \sqrt{Z_Q} 
 Q_{m(k)} \, ,
&&~~
 \hat{\bar{Q}}_{m(k)} =  
 \sqrt{Z_Q} 
 \bar{Q}_{m(k)} \, .
\label{eq:hatQ}
\eeq
The theory exits from the conformal regime at the energy scale of $M_c \sim \lambda \langle|\hat{\Phi}|\rangle \sim \bar{\lambda} \langle|\hat{\bar{\Phi}}|\rangle$.
Therefore, the $U(1)_{\rm PQ}$ breaking scale is given by 
\beq
f_{\rm PQ} &\equiv& 
\sqrt{\langle\hat{\Phi} \hat{\bar{\Phi}}\rangle} \simeq \frac{M_c}{\sqrt{\lambda\bar{\lambda}}}\nonumber\\
&=& \lmk\frac{\Lambda}{\Lambda'}\rmk^{\frac{(2+\epsilon)(1-\epsilon)}{3\epsilon}} \frac{\Lambda}{\sqrt{\lambda\bar{\lambda}}} \lmk \frac{2}{2+\epsilon}\frac{M_\Phi}{N\lambda\bar{\lambda}\Lambda} \rmk^{\frac{2+\epsilon}{3\epsilon}}.~~
\label{PQscale}
\eeq
The conformal entering scale $\Lambda$ is determined by solving RG equations for the $SU(N)$ gauge coupling $g$ and
$\lambda, \bar{\lambda}$ with initial conditions at a UV scale. 
On the other hand, $\Lambda'$ is defined as the holomorphic dynamical scale. 
These scales are closely related to each other but can be different by a factor of order unity. 
Hereafter we simply assume $\Lambda = \Lambda'$. 
Since we consider $\abs{\epsilon} \ll 1$, the last parenthesis in Eq.~\eqref{PQscale} has a large power and
one naturally obtains an intermediate scale for the spontaneous $U(1)_{\rm PQ}$ breaking
even from nearly Planck-scale values of $\Lambda$ and $M_\Phi$.

\begin{figure}[!t]
\includegraphics[width=8cm]{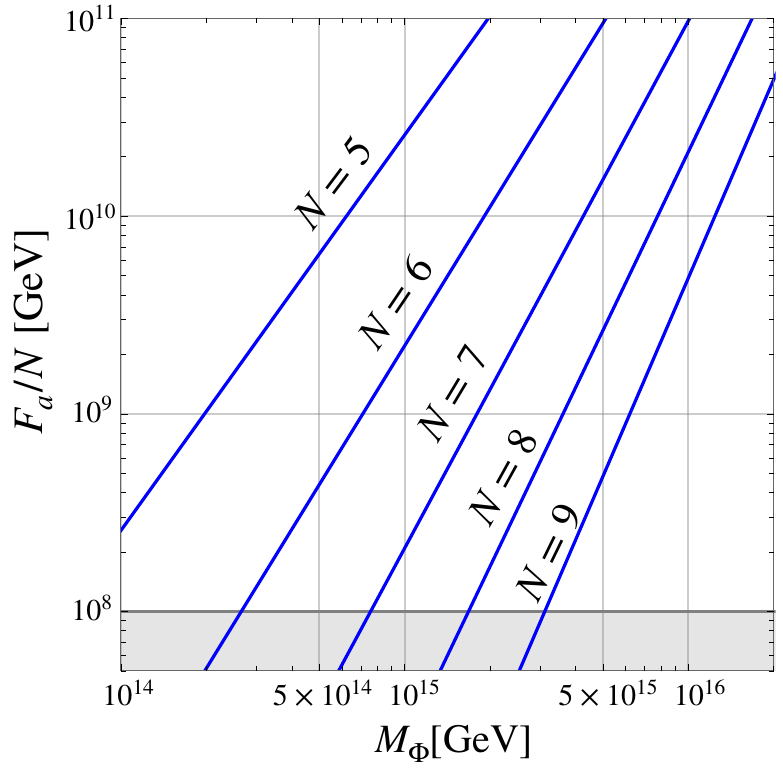}
\centering
\caption{
The axion decay constant $F_a/N$ as a function of $M_{\Phi}$ for the case of $\Lambda=0.1\Mpl$. Here we take $N=5,6,7,8,9$ from top to bottom with the lowest value of $\epsilon=2/N$. The gray shaded region is excluded by the astrophysical constraint.}
\label{fig:PQ scale}
\end{figure}

In the vicinity of the potential minimum to break the $U(1)_{\rm PQ}$,
we parameterize $\Phi$ and $\bar{\Phi}$ in terms of the axion $a$ and the saxion $\sigma$ as
\beq
\hat{\Phi} = v \exp\lmk\frac{\sigma+ia}{F_a}\rmk, \qquad 
\hat{\bar{\Phi}} = \bar{v} \exp\lmk-\frac{\sigma+ia}{F_a}\rmk,
\label{parameterize}
\eeq
where $F_a/\sqrt{2} \equiv \sqrt{v^2+\bar{v}^2}$ with $v \equiv \langle |\hat{\Phi}| \rangle$
and $\bar{v} \equiv \langle |\hat{\bar{\Phi}}| \rangle$. 
The quark supermultiplets $Q$ and $\bar{Q}$ are decoupled at the energy scale around their masses,
and then due to the QCD anomaly, the effective theory contains the axion coupling with gluons,
\beq
\Lag_{ag} =N\frac{g_s^2}{32\pi^2}\frac{a}{F_a}G_{\mu\nu}^a\tilde{G}^{a\mu\nu}.
\eeq
Here, $g_s$ denotes the $SU(3)_C$ gauge coupling, $G_{\mu\nu}^a \, (a = 1, ..., 8)$ is the gluon field strength
and $\tilde{G}^{a\mu\nu}$ is its dual.
In the conventional nomenclature, we call $F_a/N$ the axion decay constant. 
Using Eq.~\eqref{PQscale} and assuming $v=\bar{v}$, 
the axion decay constant is given as a function of $M_\Phi$,
as shown in \FIG{fig:PQ scale} 
for the cases of $N = 5, 6, 7, 8, 9$ with $\epsilon=2/N$.
We take $\Lambda = 0.1 \Mpl$. 
The values of $\lambda$ and $\bar{\lambda}$ are estimated at the conformal fixed point for each $N$, and we determine them at the two-loop level by using SARAH \cite{staub2012sarah} where $g_s$ is turned off. 
The gray shaded region is excluded by astrophysical bounds on the axion decay constant~\cite{Leinson:2014ioa,Hamaguchi:2018oqw,Leinson:2019cqv,Buschmann:2021juv,Mayle:1987as,Raffelt:1987yt,Turner:1987by,Chang:2018rso,Carenza:2019pxu}.
The figure shows that an intermediate scale is naturally generated from mass scales near the Planck scale.

%%%%%%%%%%%%%%%%%%%%%%%%%%%%%%%%%%%%%%%%%%%%%%%%%%%%%%%%%%%
%%%%%%%%%%%%%%%%%%%%%%%%%%%%%%%%%%%%%%%%%%%%%%%%%%%%%%%%%%%
\section{SUSY breaking
\label{sec:CW}}
The supersymmetric potential that we have considered so far does not stabilize the saxion direction and 
we need to take account of SUSY breaking effects. 
The soft SUSY breaking terms for our (canonically normalized) scalars are given by 
\beq
\mathcal{L}_{\rm soft} = 
&-&\sum_{m}\left( \hat{\tilde{Q}}_{m}^{*}m^{2}_{Q_{m}}\hat{\tilde{Q}}_{m} + \hat{\tilde{\bar{Q}}}_{m}m^{2}_{\bar{Q}_{m}}\hat{\tilde{\bar{Q}}}^{*}_{m}\right)\nonumber\\
&-& \sum_{k}\left(\hat{\tilde{Q}}_{k}^{*}m^{2}_{Q_{k}}\hat{\tilde{Q}}_{k} + \hat{\tilde{\bar{Q}}}_{k}m^{2}_{\bar{Q}_{k}}\hat{\tilde{\bar{Q}}}^{*}_{k}\right)
\nonumber\\
&-&m^2_{\Phi}|\hat{\Phi}|^2
-m^2_{\bar{\Phi}}|\hat{\bar{\Phi}}|^2
-\left( b\hat{\Phi}\hat{\bar{\Phi}} + {\rm h.c.} \right)  ,
\label{soft}
\eeq
where a tilde is put to represent the scalar components of $Q$ and $\bar{Q}$
and $A$-terms corresponding to the superpotential \eqref{WQ} are ignored.
In the present paper, we do not specify a SUSY breaking mechanism but just assume
all soft mass parameters are at around the TeV scale, which is much smaller than the $U(1)_{\rm PQ}$ breaking scale. 
In addition to the soft SUSY breaking terms,
the one-loop Coleman-Weinberg potential
\cite{Coleman:1973jx}
gives an important effect on the saxion direction. 
For simplicity, we assume $m_{Q_m}^2 =  m_{\bar{Q}_m}^2 = m_{Q_k}^2 = m_{\bar{Q}_k}^2 \equiv\Delta m^2$,
and then by using Eq.~\eqref{WQ} and Eq.~\eqref{soft}, the Coleman-Weinberg potential is calculated as
\beq   
&&V_{\rm CW}(\hat{\Phi},\hat{\bar{\Phi}})  \nonumber \\[1ex]
&&= 
\frac{(2+\epsilon)N^2}{64\pi^2} 
\Biggl[\lmk\lambda^2|\hat{\Phi}|^2 + \Delta m^2\rmk^2 
\log\lmk 1+\frac{\Delta m^2}{\lambda^2|\hat{\Phi}|^2}\rmk \nonumber \\
&&\quad + \lmk\bar{\lambda}^2|\hat{\bar{\Phi}}|^2+\Delta m^2\rmk^2 
\log\lmk 1+\frac{\Delta m^2}{\bar{\lambda}^2|\hat{\bar{\Phi}}|^2}\rmk\Biggr].
\label{CW}
\eeq
Summing up all contributions to the scalar potential of $\hat{\Phi}$ and $\hat{\bar{\Phi}}$, we obtain 
\beq
&&V(\hat{\Phi}, \hat{\bar{\Phi}}) = V_F + V_{\rm soft} + V_{\rm CW} \, ,
\eeq
where the first term of the right hand side is given by Eq.~\eqref{VF}.

Since the directions stabilized by the supersymmetric potential $V_F$ are much heavier than the soft mass scale, 
we can utilize $\hat{\Phi}\hat{\bar{\Phi}}=f_{\rm PQ}^2$ and focus on the flat direction. 
Then, the SUSY breaking effects completely determine VEVs for $\hat{\Phi}$ and $\hat{\bar{\Phi}}$ such as
\beq
v &\equiv& \langle |\hat{\Phi}| \rangle \nonumber\\
&\simeq& f_{\rm PQ}\lmk\frac{(3(2+\epsilon)N^2\lambda^2/64\pi^2)\Delta m^2 + m_{\Phi}^2}{(3(2+\epsilon)N^2\bar{\lambda}^2/64\pi^2)\Delta m^2 + m_{\bar{\Phi}}^2}\rmk^{1/4},~~~\label{VEV}\\
\bar{v} &\equiv& \langle |\hat{\bar{\Phi}}| \rangle \nonumber\\
&\simeq& f_{\rm PQ} \lmk\frac{(3(2+\epsilon)N^2\bar{\lambda}^2/64\pi^2)\Delta m^2 + m_{\bar{\Phi}}^2}{(3(2+\epsilon)N^2\lambda^2/64\pi^2)\Delta m^2 + m_{\Phi}^2}\rmk^{1/4}.~~~
\label{VEV2}
\eeq
For $m_\Phi^2 = m_{\bar{\Phi}}^2$ and $\lambda = \bar{\lambda}$ which lead to $v=\bar{v}=f_{\rm PQ}$,
we can simply write the saxion mass as 
\beq
m_\sigma^2 
\simeq 2m_\Phi^2+\frac{(2+\epsilon)N^2\lambda^2}{32\pi^2}\Delta m^2,
\eeq
where we have used $\Delta m^2\ll f_{\rm PQ}^2$.
Note that 
the saxion direction is stabilized only if the following condition is satisfied: 
\beq
m_\Phi^2 + \frac{(2+\epsilon)N^2\lambda^2}{64\pi^2}\Delta m^2 > 0 \, .
\label{condition}
\eeq
This condition puts a constraint on initial values of the soft SUSY breaking parameters at a UV scale  
which evolve down to the $U(1)_{\rm PQ}$ breaking scale $f_{\rm PQ}$ by RG equations.

\begin{figure}[!t]\centering
\includegraphics[width=10cm]{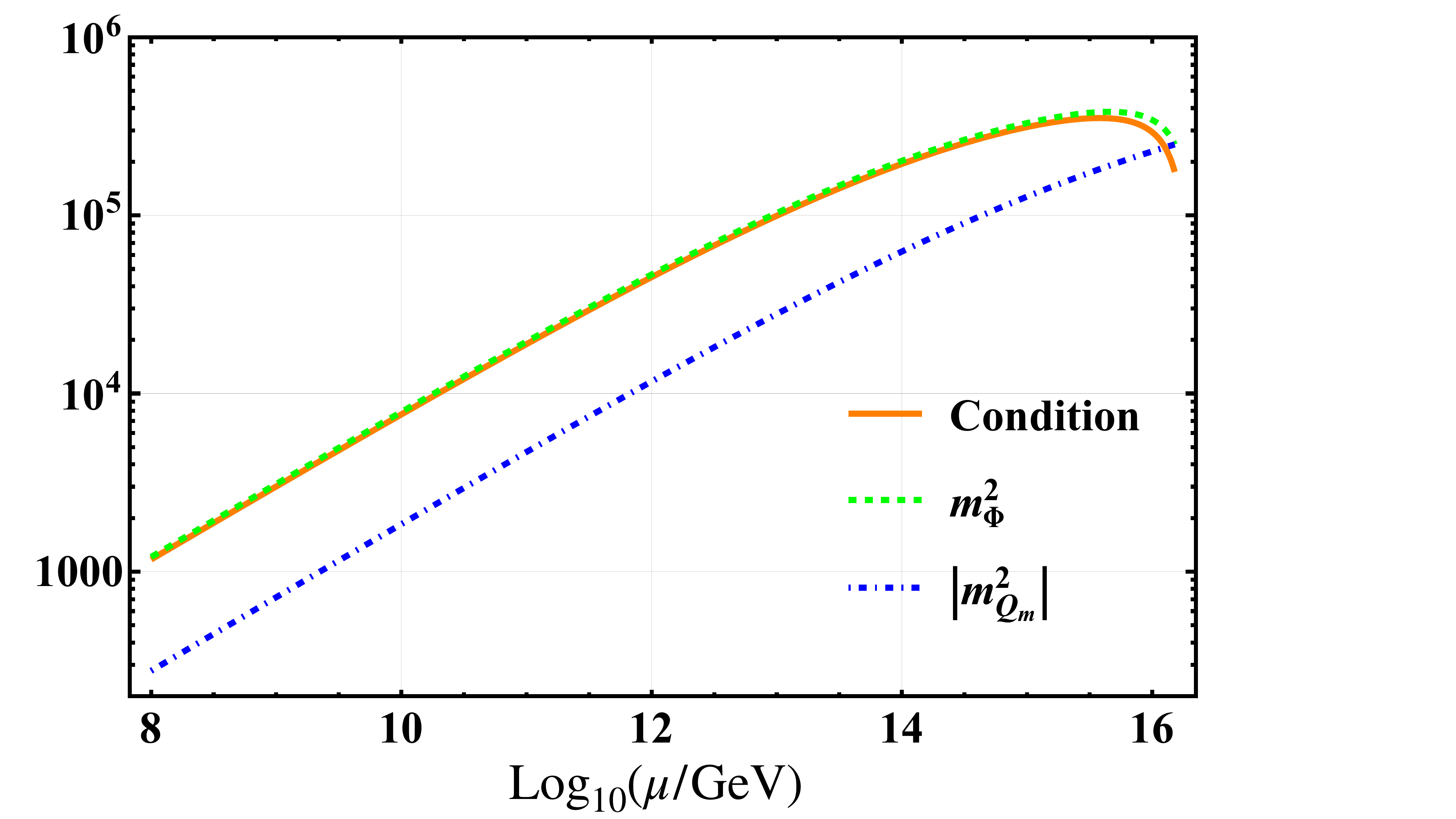}
\caption{
RG evolutions of the soft scalar masses with boundary conditions imposed at the GUT scale
$M_{\rm GUT} = 1.5\times10^{16} \, \rm GeV$. The values of them are in units of [GeV$^2$]. For symmetric initial values, the evolutions of $m_{\bar{Q}_m}^2,m_{Q_k}^2 $ and $ m_{\bar{Q}_k}^2$ are
the same as $m_{Q_m}^2$.
The orange solid curve represents the left-hand side of \eq{condition}.
}
\label{fig:sigeff}
\end{figure}

\FIG{fig:sigeff} shows an example of RG evolution of soft masses at the two-loop level calculated by using SARAH
for the case where $N=6$, $\epsilon=2/N$ and the $SU(3)_C$ effect is turned off. 
We take initial conditions at the GUT scale $M_{\rm GUT} = 1.5\times 10^{16} \, \rm GeV$ as
$g=2$, $\lambda=\bar{\lambda}=1.5$, 
$m_\Phi^2 = m_{\bar{\Phi}}^2 = (500 \, \rm GeV)^2$ and $m_{Q}^2 = -(500 \, \rm GeV)^2$. 
We also set the initial value of the $SU(N)$ gaugino mass to zero.
For simpler illustration, the absolute value of the tachyonic squark mass is taken in the figure. 
The orange solid curve represents the left-hand side of \eq{condition}. 
One can see that the condition is satisfied around and above the $U(1)_{\rm PQ}$ breaking scale. 
It also shows $m_\sigma\sim \sqrt{m_\Phi^2} = \mathcal{O}(100) \, {\rm GeV}$. 
Note that we have taken a tachyonic initial condition, $m_Q^2 < 0$, to help $m_\Phi^2$
to satisfy the condition \eqref{condition}.

Let us comment on the initial condition for the soft terms in light of a concrete SUSY breaking model.
In \FIG{fig:sigeff}, we implicitly assume gravity-mediated SUSY breaking with a somewhat heavy gravitino ($m_{3/2}\gtrsim \mathcal{O}(1)\TeV$). To suppress the gaugino mass, we can consider the minimal split SUSY model/pure gravity mediation model \cite{Giudice:1998xp,Ibe:2011aa,Ibe:2012hu,Randall:1998uk,Arvanitaki:2012ps,Arkani-Hamed:2012fhg}. 
Moreover, if the gravitino is lighter than the TeV scale, gauge-mediated SUSY breaking can be applied to our model (for a review, see e.g. ref.~\cite{Giudice:1998bp}).
On the one hand, if the messenger scale is higher than the PQ breaking scale $f_{\rm PQ}$,
the gaugino mass needs to be suppressed e.g. via direct transmission of SUSY breaking to the SM sector
\cite{Izawa:1997gs}.
On the other hand, for a lower messenger scale, the saxion can acquire a nonzero mass from the other effects, e.g. coupling to messenger fields while the detailed discussion will be left for a future exploration.

%%%%%%%%%%%%%%%%%%%%%%%%%%%%%%%%%%%%%%%%%%%%%%%%%%%%%%%%%%%
%%%%%%%%%%%%%%%%%%%%%%%%%%%%%%%%%%%%%%%%%%%%%%%%%%%%%%%%%%%
\section{$U(1)_{\rm PQ}$ quality
\label{sec:PQquality}}
Let us now consider Planck-suppressed operators that explicitly break the $U(1)_{\rm PQ}$ symmetry 
and see if the quality of the $U(1)_{\rm PQ}$ symmetry is ensured in our model without fine tuning. 
As we impose the anomaly-free discrete $Z_N$ symmetry, 
the most relevant higher dimensional operators are given by the following superpotential terms:
\beq
W_{\cancel{\rm PQ}} &=& c_{\cancel{\rm PQ}} \frac{\Phi^N}{\Mpl^{N-3}} +\bar{c}_{\cancel{\rm PQ}} \frac{\bar{\Phi}^N}{\Mpl^{N-3}}\nonumber\\[1ex]
&=& 
Z_\Phi^{-N/2}
\frac{c_{\cancel{\rm PQ}} \hat{\Phi}^N+\bar{c}_{\cancel{\rm PQ}}\hat{\bar{\Phi}}^N}{\Mpl^{N-3}} \, ,
\eeq
where we have used Eq.~(\ref{eq:hatPhi}) in the second equality. 
With the constant superpotential term, $W = m_{3/2} \Mpl^2$ where $m_{3/2}$ is the gravitino mass,
an additional axion potential is generated in supergravity, 
\beq
V_{\cancel{\rm PQ}} = \left(\frac{M_c}{\Lambda}\right)^{N\gamma_\Phi/2}\frac{\kappa_{\cancel{\rm PQ}}m_{3/2}v^N}{\Mpl^{N-3}}\cos\left(N\frac{a}{F_a}+\varphi\right).
\label{PQbreaking}
\eeq
Here,
$\kappa_{\cancel{\rm PQ}}$ is an $\mathcal{O}(1)$ constant and $\varphi$ represents a phase. 
We have assumed $v\simeq\bar{v}$ for simplicity. 
Compared with conventional SUSY axion models, 
the additional axion potential is suppressed by a factor of $\left({M_c} / {\Lambda}\right)^{N\gamma_\Phi/2}$
and the quality of the $U(1)_{\rm PQ}$ symmetry is expected to be improved in our model.

The potential \eqref{PQbreaking} should be compared with the axion potential generated from the ordinary QCD effect,
\beq
V_{\rm QCD} = \chi_0 \left[1-\cos\left(N\frac{a}{F_a}\right)\right],
\label{QCDpot}
\eeq
where $\chi_0=(75.5\MeV)^4$ is the topological susceptibility at the zero temperature \cite{GrillidiCortona:2015jxo}.
The CP symmetry is respected at the minimum of
$V_{\rm QCD}$~\cite{Peccei:1977hh,Weinberg:1977ma,Wilczek:1977pj}. 
To quantitatively estimate the quality of the $U(1)_{\rm PQ}$ symmetry, we then introduce a quality factor defined by 
\beq
\mathcal{Q}&\equiv& \frac{\left|V_{\rm \cancel{\rm PQ}}\right|_{\rm max}}{\left|V_{\rm QCD}\right|_{\rm max}} \nonumber\\
&\simeq& \left(\frac{N}{2}\frac{F_a/N}{\Lambda}\right)^{N(\frac{\gamma_\Phi}{2}+1)}\frac{(\lambda\bar{\lambda})^{N\gamma_\Phi/4}\kappa_{\cancel{\rm PQ}}m_{3/2}\Lambda^N}{\Mpl^{N-3}\chi_0},~~~~~
\eeq
where the subscript ``max" represents the maximum height of a potential.
Without fine tuning in the phase, $\varphi = \mathcal{O}(1)$, 
the potential minimum of the axion is shifted by a factor of $\mathcal{Q}$ from the CP-conserving minimum. 
Therefore, $\mathcal{Q}$ must be smaller than $10^{-10}$ to address the strong CP problem without fine tuning.

\begin{figure}[!t]
\includegraphics[width=8.5cm]{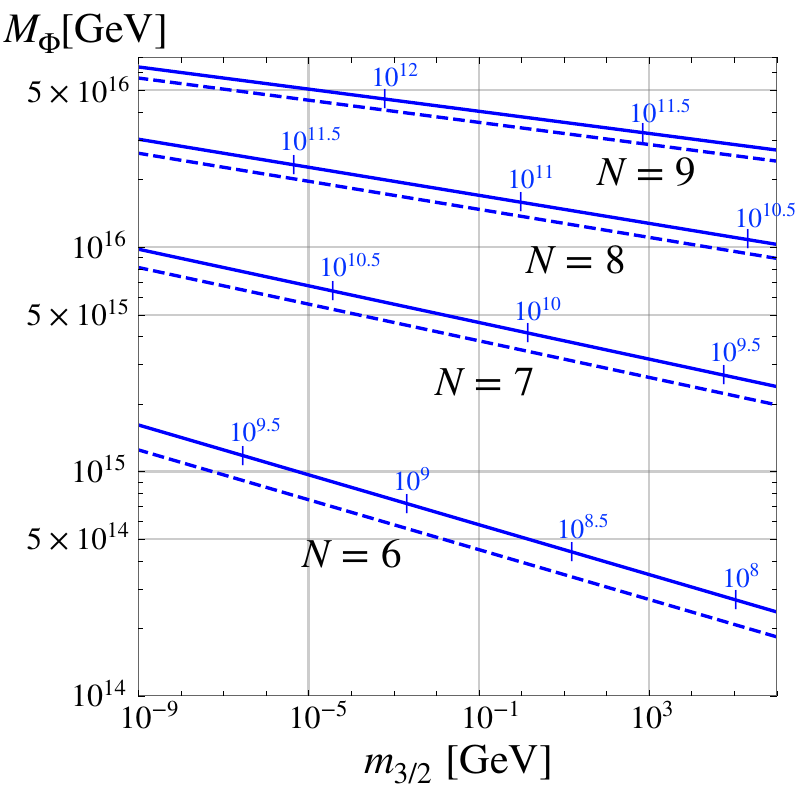}
\centering
\caption{
Contours of the quality factor for $\mathcal{Q}=10^{-10}$ (solid line) and $\mathcal{Q}=10^{-12}$ (dashed line)
in the $M_{\Phi}\,\text{-}\,m_{3/2}$ plane. 
We take $\Lambda=0.1\Mpl$ and $\kappa_{\rm \cancel{\rm PQ}}=1$.
The scales on each line denote the values of the axion decay constant $F_a/N \, [{\rm GeV}]$.
}
\label{fig:Mph}
\end{figure}

\FIG{fig:Mph} shows contour plots of the quality factor $\mathcal{Q}$ in the $M_\Phi\,\text{-}\,m_{3/2}$ plane
for $N=6,7,8,9$ with $\epsilon=2/N$. 
The solid and dashed lines correspond to the contours for $\mathcal{Q}=10^{-10}$ and $10^{-12}$, respectively, 
where we take $\kappa_{\cancel{\rm PQ}}=1$ and $\Lambda=0.1\Mpl$. 
The scales on each line denote the values of the axion decay constant $F_a/N$ in the unit of GeV.
One can see from the figure that the quality problem can be addressed or relaxed even for a relatively small $N$.

%%%%%%%%%%%%%%%%%%%%%%%%%%%%%%%%%%%%%%%%%%%%%%%%%%%%%%%%%%%
%%%%%%%%%%%%%%%%%%%%%%%%%%%%%%%%%%%%%%%%%%%%%%%%%%%%%%%%%%%
\section{Discussions
\label{sec:conclusion}}

We have explored a dynamical mechanism to realize the emergence of the $U(1)_{\rm PQ}$ symmetry
and its spontaneous breaking at an intermediate scale for an axion solution to the strong CP problem.
Our approach was to consider a new SQCD sector near the middle of conformal window
that couples to fields spontaneously breaking the $U(1)_{\rm PQ}$ symmetry.
It was shown that a large anomalous dimension of the $U(1)_{\rm PQ}$ breaking fields leads to
the suppression of explicit $U(1)_{\rm PQ}$-violating higher dimensional operators.
The supersymmetric mass term of the $U(1)_{\rm PQ}$ breaking fields gives a small marginally relevant deformation
in the superpotential, and it is finally balanced with a non-perturbative effect of the SQCD
to generate a desired $U(1)_{\rm PQ}$ breaking vacuum.
Our scalegenesis mechanism is not limited to the generation of the $U(1)_{\rm PQ}$ breaking scale, but
one may be able to consider a wide range of applications.

In \SEC{sec:CW}, we discussed the stabilization of the saxion direction by SUSY breaking effects
and found that the saxion mass is of the order of the soft mass scale $(\simeq\mathcal{O}(100) \, \rm GeV)$. 
As in the case of the axion, the saxion is initially misaligned from its potential minimum during inflation. When the Hubble parameter becomes comparable to the mass, it begins to oscillate and behaves like a non-relativistic matter. 
In our model, the saxion dominantly decays into gluons at the rate of\footnote{
Expanding the kinetic term of the saxion, we can find another decay mode into two axions,
but for $v \simeq \bar{v}$, it is significantly suppressed.} 
\beq
\Gamma_{\sigma\rightarrow gg} =\frac{\alpha_s^2}{16\pi^3}\frac{m_\sigma^3}{F_a^2} \, 
\eeq
with $\alpha_s\equiv g_s^2/4\pi$.
For our focused range of $F_a$, the saxion decays much before the Big Bang Nucleosynthesis
and is cosmologically harmless.

In our scenario, the $U(1)_{\rm PQ}$ symmetry must be spontaneously broken during inflation.
Otherwise, domain walls dominate the Universe
because the domain wall number is $N>1$.
One still has to care about the restoration of the $U(1)_{\rm PQ}$ symmetry after inflation
due to thermal corrections to the $U(1)_{\rm PQ}$ breaking fields.
However, our new (s)quarks have masses of $\mathcal{O}(f_{\rm PQ})$,
and moreover the potential overproduction problem of axinos and gravitinos,
which is mentioned in App.~\ref{sec:axino},
indicates the reheating temperature much lower than $f_{\rm PQ}$.
Thus we expect that the (s)quarks are decoupled so that the $U(1)_{\rm PQ}$ symmetry remains broken.

%%%%%%%%%%%%%%%%%%%%%%%%%%%%%%%%%%%%%%%%%%%%%%%%%%%%%%%%%%%
%%%%%%%%%%%%%%%%%%%%%%%%%%%%%%%%%%%%%%%%%%%%%%%%%%%%%%%%%%%
\section*{Acknowledgments}
We thank Ryosuke Sato, Yoshihiro Shigekami and Motoo Suzuki for useful discussions.
YN is supported by the Natural Science Foundation of China under grant No. 12150610465.
The work is partly supported by 
the Graduate Program on Physics for the Universe of Tohoku University (SN), JST SPRING, Grant Number JPMJSP2114 (SN), 
JSPS KAKENHI Grant Numbers 20H05851 (MY) and 23K13092 (MY). 
MY was also supported by MEXT Leading Initiative for Excellent Young Researchers.

\appendix

\section{Fermion components of the $U(1)_{\rm PQ}$ breaking fields
\label{sec:axino}}

We here discuss the fermion components of the $U(1)_{\rm PQ}$ breaking fields $\Phi, \bar{\Phi}$.
In the SUSY limit, they have mass terms as 
\beq
\Lag &\supset& -\frac{1}{2} \sum_{i,j} \lmk \frac{\del^2 W}{\del \Phi_i\del\Phi_j}\psi_i\psi_j + {\rm h.c.} \rmk
\nonumber\\[1ex]
&\equiv& -\Psi^T M\Psi \, ,
\eeq
where $\Phi_i$ takes $\Phi$ and $\bar{\Phi}$, $\psi_i$ denotes the corresponding Weyl fermion,
and we define $\Psi=(\psi, \bar{\psi}, \psi^*, \bar{\psi}^*)^T$.
Taking account of Eqs.~\eqref{gaugino}, \eqref{PhibarPhi},
the mass matrix is given by
\beq
M = \left(
  \begin{array}{cc}
    M_1 & 0 \\
     0  & M_2 \\
  \end{array}
\right),
\eeq
in terms of $2 \times 2$ matrices,
\beq
M_1 &=& \frac{\epsilon}{4} M_\Phi \lmk
  \begin{array}{cc}

    \Phi/\bar{\Phi} & 1\\
    1 & \bar{\Phi}/\Phi \\
  \end{array}
\rmk,\\[1ex]
M_2 &=& \frac{\epsilon}{4} M_\Phi \lmk
  \begin{array}{cc}
    \Phi^*/\bar{\Phi}^* & 1\\
    1 & \bar{\Phi}^*/\Phi^* \\
  \end{array}
\rmk.
\eeq
Here, we have used \EQ{flatdir}. 
Eigenvalues for the mass matrix $M$ are then given by 
\beq
\frac{\epsilon}{4}\lmk \frac{\Phi}{\bar{\Phi}} +\frac{\bar{\Phi}}{\Phi} \rmk M_\Phi,
~~~\frac{\epsilon}{4}\lmk \frac{\Phi^*}{\bar{\Phi}^*} +\frac{\bar{\Phi}^*}{\Phi^*} \rmk M_\Phi, 
~~~0, 
~~~0,
\eeq
which correspond to a massive fermion and a massless fermion.
As the axion and the saxion are massless in the SUSY limit, the axino exists
as a two-component massless Weyl fermion.\footnote{We can include
the $U(1)_{\rm PQ}$-violating Planck-suppressed superpotential terms,
but their contributions are negligible compared to SUSY breaking effects.
}
We expect that the axino acquires a mass of around or larger than the gravitino mass
by coupling the $U(1)_{\rm PQ}$ breaking fields to a SUSY breaking sector, e.g. through supergravity effects. 
The axino mass is then model-dependent.
For example, one may consider the supergravity effect,
\begin{equation}
\begin{split}
\Lag_{\rm axino} &= \int d^4\theta \frac{(A+A^\dagger)^2(X+X^\dagger)}{\Mpl} \\
&\sim -\frac{1}{2}m_{3/2}\tilde{a}\tilde{a} \, ,
\end{split}
\end{equation}
where $X$ and $A$ denote a SUSY breaking field and the axion chiral supermultiplet, respectively.
The axino mass is of the order of the gravitino mass.

The overproduction of axinos gives a constraint on the reheating temperature depending on the axino mass. 
Since the gravitino has a similar upper bound, the reheating temperature is stringently constrained
in the whole range of the gravitino mass
\cite{Cheng:2001ys}.
Our model also requires such a low reheating temperature.

\bibliography{reference}

\end{document}